\documentclass[aps,prl,reprint]{revtex4-1}

\usepackage{graphics}      
\usepackage{graphicx}      
\usepackage{longtable}     
\usepackage{url}           
\usepackage{bm,color}      
\usepackage{amssymb, amsmath, amsthm, enumerate, epsfig,subfigure,setspace,ulem}
\usepackage{multirow}
\theoremstyle{plain}
\newtheorem*{prop*}{Proposition}

\definecolor{red}{rgb}{0.8, 0.25, 0.33}
\definecolor{green}{rgb}{0.0, 0.5, 0.0}

\begin{document}

\title{Macroscopic noise amplification by asymmetric dyads in non-Hermitian optical systems for generative diffusion models}

\author{Alexander Johnston and Natalia G. Berloff}

\email[correspondence address: ]{N.G.Berloff@damtp.cam.ac.uk}

\affiliation{Department of Applied Mathematics and Theoretical Physics, University of Cambridge, Cambridge CB3 0WA, United Kingdom}

\date{\today}

\begin{abstract}
A new generation of sensors, hardware random number generators, and quantum and classical signal detectors are exploiting strong responses to external perturbations or system noise. Here, we study noise amplification by asymmetric dyads in freely expanding non-Hermitian optical systems. We show that modifications of the pumping strengths can counteract bias from natural imperfections of the system’s hardware, while couplings between dyads lead to systems with non-uniform statistical distributions. Our results suggest that asymmetric non-Hermitian dyads are promising candidates for efficient sensors and ultra-fast random number generators. We propose that  the integrated light emission from such asymmetric dyads  can be efficiently used  for an  analog all-optical degenerative diffusion models of  machine learning to overcome the digital  limitations of such models in processing speed and energy consumption.
\end{abstract}

\maketitle
{\it Introduction}. For open quantum and classical systems, it can be justified to consider effective Hamiltonians that are non-Hermitian or non-self-adjoint. The non-Hermiticity of the Hamiltonian implies that energy eigenvalues are generally complex numbers. In classical optical systems, where resonant frequencies and modes represent energy eigenvalues and eigenstates, non-Hermiticity arises from gain and dissipation. Non-Hermitian systems driven by gain and dissipation provide a versatile platform for exploring pattern-forming  in mechanical and electronic systems, photonics,  optics, atomic systems,   optomechanics, fluids, biological transport, and acoustics \cite{stockmann2002effective, ruter2010observation, wiersig2011structure}. 
The competition between conservative and non-conservative processes in photonics has led to disparate artificial structures such  as optical fibres \cite{knight1998photonic}, photonic crystals \cite{joannopoulos1997photonic},  metamaterials \cite{shalaev2007optical}, and Bose-Einstein condensates formed from either exciton-polaritons \cite{Kasprzak2006} or photons \cite{klaers2010bose}. Such non-Hermitian systems display a range of unexpected and unique behaviours \cite{el2018non}, from topological energy transfer \cite{xu2016topological, ghosh2016exceptional} and single-mode lasing \cite{hodaei2014parity, miri2012large, feng2014single} to robust biological transport \cite{nelson1998non}. A dramatic manifestation of non-Hermitian physics is the macroscopic response to small perturbations, e.g., global spiral wave asymmetry in Belousov-Zhabotinsky chemical reactions \cite{muller1995spiral}, patterns  induced by thermal fluctuations  just below convection onset in fluids \cite{rehberg1991thermally}, or the global direction of vortex rotation or wave chirality in disparate systems governed by the complex Ginzburg-Landau equations \cite{staliunas2003transverse, keeling2008spontaneous,alperin2021multiply}.

Another practical use of noise-sensitive amplification lies in (true) hardware random number generation (hRNG) where system noise should be statistically random, sampled sufficiently fast, and able to be macroscopically amplified to a measurable level and suitably processed. In contrast, pseudo-random number generation is implemented by a computer algorithm. Thermal or quantum noise, the photoelectric effect, involving a beam splitter, and other quantum phenomena (e.g., shot noise \cite{shen2010practical}, nuclear decay \cite{herrero2017quantum}, and spontaneous parametric down-conversion \cite{marandi2012all}) can all generate low-level, statistically random  signals used for hRNG. Random numbers are obtained after  randomly varying hardware noise is  repeatedly sampled with the system-dependent output data rate. True, efficient, and fast random number generation is crucial for a variety of industries, from cryptography and finance  \cite{petrie2000noise,yoshimura2012secure, koizumi2013information, honjo2009differential} to large-scale parallel computation \cite{miyazawa2009implementation, rousseau2008random, katzgraber2010random}. It is, therefore, essential to search for novel hardware systems that have non-trivial, statistically controllable and easily detectable  macroscopic reactions to background noise while providing an ultra-fast output rate.
The intensive  use of  digital noise generation is expected in diffusion models in machine learning, which have recently become the state-of-the art choice for image synthesis \cite{sohl2015deep, dhariwal2021diffusion, ramesh2021zero, meng2021sdedit, nichol2021glide, bansal2022cold}. A crucial part of diffusion models is the injection of Gaussian noise.
However, the digital diffusion process is computationally and energy intensive, requiring significant processing power, memory bandwidth and high throughput times, especially when dealing with high-resolution images or large datasets. Using optical analog hardware in the forward and reverse stages of the diffusion process can allow us to 
 perform computations with significantly lower energy consumption and faster times, generate less heat compared to electronic systems and avoid physical limitations of electronic circuits, such as electron mobility and heat dissipation.

In this Letter, we propose that freely expanding non-Hermitian condensates, such as microcavity exciton-polariton or photon condensates, have potential use as sensors, detectors, or hRNGs since they fulfil the necessary criteria for efficient operation: macroscopic response to small system noisy perturbations, capacity for calibration of system imperfections, fast output data rate and compatibility with all-optical transmission. These properties could supply the Gaussian noise all optically to implement diffusion models in machine learning.

Unlike conservative systems, gain-dissipative optical nonequilibrium systems can have peculiar asymmetric states resulting from completely symmetric conditions, even in an ideal system. One such state is an asymmetric dyad: two geometrically-coupled non-Hermitian condensates with macroscopic, unequal occupations, and a phase difference, despite having equal pumping intensities (and other identical conditions) \cite{johnston2021artificial}. The degree of population asymmetry  and phase difference varies depending on the losses, nonlinearities,  gain intensity and shape, and the distance between the condensates and can be made arbitrarily large or small—the asymmetry orientation forms spontaneously in response to any (even insignificant) bias in the system.

We first demonstrate that, 
starting from random ultra-low-level noisy initial conditions, simulating hardware noise, the two possible directions of the final orientation of the dyad asymmetry are equally likely. We develop an error correction scheme and demonstrate that slight modifications of the pumping strength at one condensate site can ensure both orientations are equally likely even in the presence of small asymmetries in the physical sample itself. We show how the lattice of such dyads can be used to generate the Gaussian noise using integrated light intensity and superimposed with the image to perform the forward part of diffusion process in machine learning.

\begin{figure}[h!]
     \includegraphics[width=\columnwidth]{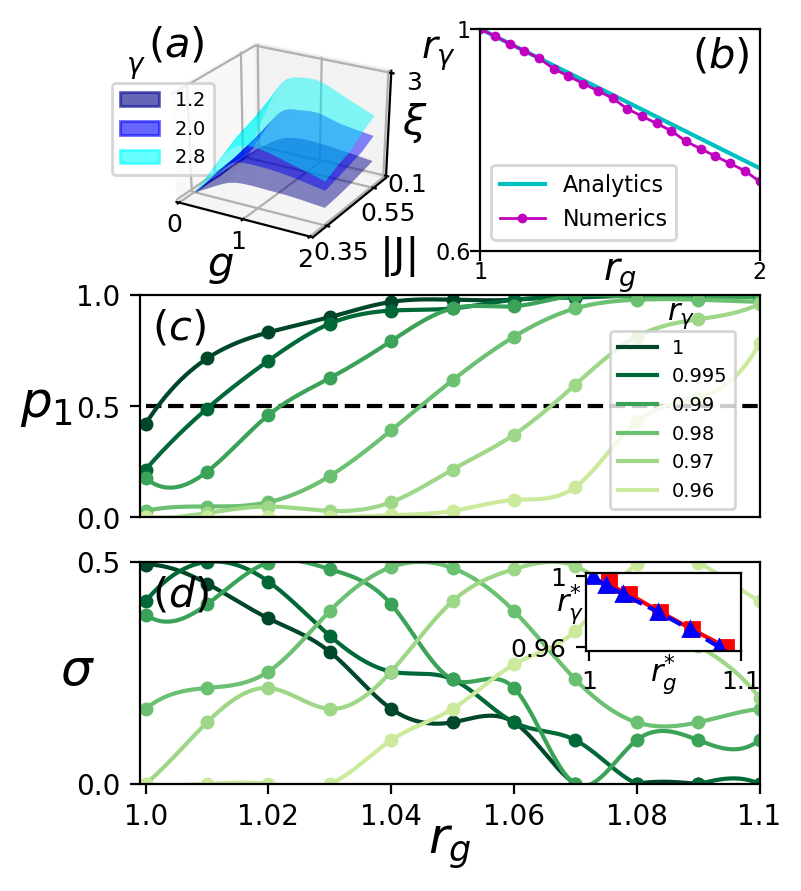}
     \caption{(a): Surfaces defining upper bounds (in terms of $\xi$) of the regions of $g-|J|-\xi$ space in which asymmetric dyads form, for three $\gamma$ values. The regions are symmetric with respect to the sign of $J$. For each value of $\gamma$, the set of $N$ equations described by Eq.~\ref{psi_i_equation} was solved for $N=2$ (with random initial conditions) for $12500$ sets of parameters within the $g-|J|-\xi$ space shown. A smooth surface was fitted at the boundary of the set of points at which asymmetric states were stable. For a dyad with equal occupations, (b) shows the relationship between the ratios $r_{g} \equiv (g + \epsilon)/g$ and $r_{\gamma} \equiv \tilde{\gamma}/\gamma^0$ required to maintain equal occupation. $r_{g}$ ($r_{\gamma}$) represents the ratio of the blueshift (pumping strength) between the two condensation sites. The analytical approximation of Eq.~\ref{gamma_1_equation} for small $\epsilon$ (i.e. $r_{g} \approx r_{\gamma} \approx 1$) is compared against numerically-calculated values. For (b), $J=0.45$, $\gamma=1.8$, $g=0.4$, and $\xi=2$. (c) and (d): The proportion of times the system converges to a ``1" state, $p_{1}$, as a function of $r_{g}$ for a variety of $r_{\gamma}$ values is shown in (c). $p_{1} \equiv n_{1}/(n_{0} + n_{1})$, where $n_{1}$ ($n_{0}$) is the total number of ``1" (``0") states. The standard deviation of the set of final states, $\sigma$, as a function of $r_{g}$ is shown in (d) for the same $r_{\gamma}$ values. Each data point was calculated by running $1000$ simulations of the set of $N$ equations described by Eq.~\ref{psi_i_equation} for $N=2$ with random initial conditions and a single coupling, $J$, between the two condensates. For each value of $r_{\gamma}$, there is a critical point $r_{g} = r_{g}^{*}$ at which $p_{1} = \sigma = 0.5$. The inset of (d) shows a plot of these critical points $(r_{\gamma}^{*},r_{g}^{*})$ determined empirically using cubic spline methods from (c) when $p_{1} = 0.5$ (blue triangles) and (d) when $\sigma = 0.5$ (red squares). 
     For (c) and (d), $J=0.55$, $\gamma=2.8$, $g=0.5$, and $\xi=5/3$. } 
    \label{Combined dyad plot}
\end{figure}

{\it Dynamics of condensate centres and asymmetric non-Hermitian dyads.} We model the photonic non-Hermitian system by the following system of $N$ equations describing a network of $N$ optically excited and interacting nonequilibrium condensate centres (CCs):
\begin{equation}
    \dot{\psi_{i}} = -i|\psi_{i}|^2\psi_{i} - \psi_{i} + (1 - ig)\Big[\Big(\frac{\gamma}{1 + \xi |\psi_{i}|^2} \Big) \psi_{i}  + \sum_{j \neq i} J_{ij}\psi_{j} \Big],
    \label{psi_i_equation}
\end{equation}
where $\psi_{i}(t) = \sqrt{\rho_{i}}\exp{[i\theta_{i}]}$ is the complex amplitude of the $i^{th}$ CC (and $\rho_{i}$ and $\theta_{i}$ are its occupation and phase, respectively), $\gamma$ is the pumping strength, $g$ is the detuning strength (blueshift), $\xi$ characterises the relative strength of the system's nonlinearities, and $J_{ij}$ is the coupling strength between the $i^{th}$ and $j^{th}$ CCs. The parameter $g$ is often referred to as the “cavity blueshift” since it provides a measure of the polariton-exciton interaction strength, which induces a blueshift in the frequency of the light emitted from the microcavity \cite{bajoni2008polariton}. Typically, $|J_{ij}|<1$ in these dimensionless units. These equations describe a variety of coupled oscillator systems with saturable nonlinearity, from nonequilibrium condensates (such as exciton-polariton or photon condensates) to lasers and non-parametric oscillators \cite{staliunas2003transverse,aleiner2012radiative,johnston2021artificial, kalinin2019polaritonic}. These equations can be  derived using the tight-binding approximation of the mean-field complex Ginzburg-Landau equation in the fast reservoir regime \cite{johnston2021artificial, kalinin2019polaritonic} or using the full mean-field Maxwell-Bloch equations for laser cavities \cite{dunlop1996generalized}. The system exhibits a range of behaviours depending on system parameters, such as evolution to a stationary state, periodic or chaotic oscillations \cite{aleiner2012radiative,johnston2021artificial}. The combination of nonlinearity, gain, and dissipation results in a region of parameter space in which the density and phase asymmetry of stationary states appears even with identical site conditions, due to quantum or classical noise amplification. Fig.~\ref{Combined dyad plot}(a) shows the region in $g-|J|-\xi$ space in which asymmetric states occur for the system described by Eq.~\ref{psi_i_equation}, for three $\gamma$ values. The three surfaces bound the region from above (in terms of $\xi$) in each case. 
 For a dyad in such a parameter regime, 
 the two possible configurations are equally likely under ideal conditions and correspond to two possible directions for the asymmetry. This asymmetry can be defined as the direction from the higher to the lower density condensate. Dyads oriented in one direction can be labelled as a ``1" state, with those in the opposing orientation labelled as a ``0" state.

 {\it Error correction}. The physical sample on which the condensates are prepared will likely have some intrinsic asymmetry, which may result in a slight preference for the higher-density component of the asymmetric dyad to condense at one site in particular. This would create a non-uniform distribution for the initial noise, leading to a biased distribution for the dyad orientations. Below we show how to counteract such intrinsic asymmetry by modifying the pumping strength applied to one of the condensates.

We model asymmetry in the physical sample by considering a small perturbation $\epsilon$ to $g$ for one of the condensates given by Eq.~\ref{psi_i_equation}. We show it is possible to modify the pumping strength of this condensate $\tilde {\gamma}$ to again achieve an equally-likely distribution of asymmetry in the dyad. The steady state satisfies:
\begin{eqnarray}
    -i\mu \psi_{1} &=& -i|\psi_{1}|^2\psi_{1} - \psi_{1} + (1 - ig) \nonumber\\
    && \quad \quad \quad \quad\times \Big[\Big(\frac{\gamma}{1 + \xi |\psi_{1}|^2} \Big) \psi_{1}  + J\psi_{2} \Big],
    \label{psi_1_equation} \\
    -i\mu\psi_{2} &=& -i|\psi_{2}|^2\psi_{2} - \psi_{2} + (1 - i(g + \epsilon))\nonumber\\
    &&\quad\quad\quad \quad\times\Big[\Big(\frac{\tilde{\gamma}}{1 + \xi |\psi_{2}|^2} \Big) \psi_{2}  + J\psi_{1} \Big],
    \label{psi_2_equation}
\end{eqnarray}
where $\mu$ is the chemical potential defined by $i\dot{\psi_i}=\mu \psi_i$ and $\psi_i$ are the new condensate wavefunctions modified from their steady-state values of Eq.~\ref{psi_i_equation}. We will consider small deviations from the unperturbed values marked by superscript $0$ by writing $\psi_1=a_1^0 + \epsilon a_1^1, \psi_2=(a_2^0 + \epsilon a_2^1) \exp(i\theta^0 + i\epsilon \theta^1)$ with the chemical potential $\mu=\mu^0 + \epsilon \mu^1$ and pumping $\tilde{\gamma}=\gamma^0+\epsilon \gamma^1 $. We then linearize Eqs.~(\ref{psi_1_equation}) and (\ref{psi_2_equation}) for small $\epsilon$. To the leading order we recover the steady states of Eq.~(\ref{psi_i_equation})
\begin{eqnarray}
-i\mu^0 &=& -i({a_j^0})^2 - 1 + (1 - i g)\biggl[\frac{\gamma^0}{1 + \xi ({a_j^0})^2} \nonumber \\
&&\quad\quad \quad \quad\quad \quad+  J \frac{a_l^0}{a_j^0}\exp(i(-1)^j\theta^0)\biggl],
\label{epsilon0}
\end{eqnarray}
where $j=1,l=2$ or $j=2,l=1$.
At the first order in small $\epsilon$,   we get four real linear equations that we can solve for $a_1^1,a_2^1,\gamma^1$ and $\theta^1$, while keeping $\mu^1$ as a free parameter. The expression for $\gamma^1$ should be invariant under the change $a_1^0 \longleftrightarrow a_2^0$, $\theta^0 \longleftrightarrow -\theta^0.$ This consideration fixes the $\mu^1$ value and shows it is  possible to modify the pumping by $\gamma^1$ to compensate for the asymmetry in $g$. Simpler analytics can be obtained by considering the limit of small
asymmetry between the condensates: $a_2^0=(1 + \delta)a_1^0,$ $\delta \ll 1.$ To the leading order in $\delta$, Eq.~(\ref{epsilon0}) gives the unperturbed solutions
$(a_1^0)^2=(\gamma^0 + |J| - 1)/(1 - |J|)\xi$, $\theta^0=0$ (for $J>0$) and $\theta^0=\pi$ (for $J<0$) and $\mu^0= g + (a_1^0)^2$. These states correspond to  the equal occupancy states. Considering the leading order expansions  in $\delta$ of the first order equations in small $\epsilon$, we get 
\begin{eqnarray}
\label{gamma_1_equation}
\gamma^1&=& - \frac{g \gamma^0}{(1 + g^2)(1 - |J|)},  \mu^1=\frac{1}{2}-\frac{\gamma^0 g}{2(1 + g^2)(1 - |J|)^2 \xi},\nonumber\\
 \theta^1&=&\frac{1}{2(1+g^2)|J|}.
\end{eqnarray}
 Fig.~\ref{Combined dyad plot}(b) shows the ratios $r_{g} \equiv (g + \epsilon)/g$ and $r_{\gamma} \equiv \tilde{\gamma}/\gamma^0$  required to maintain equal occupation. Values derived using the analytical approximation of Eq.~\ref{gamma_1_equation} are shown alongside numerically-calculated values. 
 The relationship from numerics is also approximately linear, and it converges to the analytical solution for low $\epsilon$, as expected. 

When $\delta$ is not small, we numerically investigate the effect of $r_{g}$ and $r_{\gamma}$ on the proportion of ``1" states produced, $p_{1} \equiv n_{1}/(n_{0} + n_{1})$, where $n_{1}$ ($n_{0}$) is the total number of ``1" (``0") states the system converges to after $1000$ trials. Fig.~\ref{Combined dyad plot}(c) shows a series of curves depicting this proportion as a function of $r_{g}$ for a range of $r_{\gamma}$ values, while Fig.~\ref{Combined dyad plot}(d) shows the standard deviation of the set of spin values (i.e. ${0,1}$) collected. When $
r_{g} = r_{\gamma} = 1$,  $p_{1} = \sigma = 0.5$, as expected from a binomial distribution with $p=0.5$, i.e. a fair coin toss. Increasing $r_{g}$, causes $p_{1}$ to increase towards a plateau at which all states are ``1" states. Decreasing $r_{\gamma}$, however, counteracts this effect, shifting the starting point of the curve such that $p_{1} < 0.5$ when $r_{g} = 1$. Since $p_{1}$ increases as $r_{g}$ increases, this ensures there is always an optimum pair of values $(r_{g}^{*}, r_{\gamma}^{*})$ at which $p_{1} = \sigma = 0.5$. Cubic spline methods were used to empirically determine a set of six $(r_{g}^{*}, r_{\gamma}^{*})$ values in the case when $p_{1} = 0.5$ and when $\sigma = 0.5$, respectively. An inset in Fig.~\ref{Combined dyad plot}(d) shows a plot of $r_{\gamma}^{*}$ against $r_{g}^{*}$ calculated using $p_{1} = 0.5$ intersection points (blue triangles) and $\sigma = 0.5$ intersection points (red squares). For this range of perturbed $g$, the analytics obtained to the first order in $\epsilon$ gives the correct, numerically-observed slope of $-0.448$. 
Thus, any intrinsic asymmetry in the physical sample can be counteracted by modifying the pumping strength at one site. The system can, therefore, be engineered to ensure the resultant orientation of population asymmetry behaves as a fair coin toss. A lattice of such dyads, constructed with zero coupling between condensates in different dyads, would ensure the orientation of each dyad is independent of all the others. These two facts (the system's unbiased choice of dyad asymmetry orientation and the independence of individual dyads) enable a lattice of such dyads to produce a normal distribution with a given mean and variance (as demonstrated later). It also means such a system could be used as a sensitivity device. Non-Hermitian systems have recently shown promise in exploiting exceptional points for sensing \cite{wiersig2020review}, including in microcavities \cite{richter2019voigt, liao2021experimental}. In our system, once the unbiased platform is set up, any subsequent asymmetry in the orientation statistics could be attributed to some internal defect in the sample or (if one condensate is shielded) an external effect (e.g., low-intensity radiation) which temporarily biases the results. This would allow one to map when such external effects impact the dyad. For certain dyad locations, some defects may have an equivalent impact on both condensates, yielding no change in orientation statistics. However, a change in the statistics could be detected by conducting several trials with different dyad locations (by varying the locations of laser light). 
In Supp. Inf. we also consider the chains of asymmetric dyads and the couplings between them that allow the generation of biased distributions.

{\it Gaussian noise for generative diffusion models}. Next, we consider a particular application of the lattice of polariton dyads in the generative diffusion models of machine learning. Such models, also known as score-based diffusion models \cite{sohl2015deep}, have rapidly evolved to become the leading member of deep generative models, surpassing the long-standing dominance of generative adversarial networks in image synthesis \cite{goodfellow2014generative}. Diffusion models have not only excelled in producing high-quality images \cite{ho2020denoising,song2020score}, but have also shown versatility across various domains such as audio generation \cite{liu2023audioldm}, video content creation \cite{ho2022imagen,singer2022make}, and beyond. Their applications extend to computer vision, natural language processing, temporal data modeling, multi-modal modeling, robust machine learning, and interdisciplinary fields like computational chemistry and medical image reconstruction. A diffusion model consists of two processes: a forward process that gradually transforms data into noise and a generative denoising process that reverses the effect of the forward process and learns
to transform the noise back into data. Both processes use iterative sampling that, in the forward process, slowly destroys data information by adding normal noise to the image at the
previous step until it fully replaces it with Gaussian noise. The denoising process also performs iterative sampling from a normal distribution. Still, the mean of the distribution
is derived by subtracting the noise, estimated by a neural network, from the image at the previous step while the variance is equal to the one used in the forward
process. 

 We propose that the unbiased distributions of noise generated in our scheme could generate the Gaussian noise required for diffusion models in machine learning \cite{sohl2015deep, dhariwal2021diffusion, ramesh2021zero, meng2021sdedit, nichol2021glide, bansal2022cold}. The fine control over the coupling between individual dyads means that a square lattice of independent dyads can be created that can map to a set of pixels. The Gaussian noise could then be generated as follows. Initially, the number of pixels required to adequately represent the granularity of the desired noise distribution must be determined. If the desired distribution has a broad dynamic range, multiple pixels can be used to represent each sample. If we have $n$ dyads (pixels), each firing $+1$ or $-1$ with equal probabilities, the sum of their values will follow a binomial distribution. By scaling and shifting appropriately, this can be made to approximate a standard normal distribution (by the Central Limit Theorem) for large $n$. This allows for a transformation from binary noise to a (approximately) Gaussian distribution. The Gaussian distribution can then be mapped into a uniform distribution (or be used as is). The technique of inverse transform sampling can then be used to generate samples from any desired distribution. Given a continuous cumulative distribution function $F(x)$ for the desired distribution, one can apply $F^{-1}(x)$ to samples from a uniform distribution to obtain samples from the desired distribution. 
The transformed noise from our analogue dyad platform could be used to replace or augment the noise term. However, in this scheme, many digital readouts of the individual dyads are required for a single step of the noise injection. Instead, in the case of polariton dyads, we can use the optical transmission of the integrated intensities of polariton lattices. Therefore, we can dramatically reduce the number of digital conversions or eliminate them in favour of all-optical diffusion processes.  

Polariton condensates have a coherence lifetime that depends on the quality of the microcavity, polariton lifetime, temperature, strength of polariton-polariton interactions, pump power, etc. and can range from a few picoseconds to several tens or hundreds of picoseconds \cite{love2008intrinsic, whittaker2009coherence, orfanakis2021ultralong}. A short coherence time implies that the condensates are created and destroyed during measurements that give rise to the integrated intensities (II) being observed $I=|\sum_{i=1}^{n}\psi_i|^2/n^2,$ where $n$ is set by the time of the measurement divided by the coherence time. Let's consider a single polariton dyad with the equally likely states $(q_1,q_2)$ and $(q_2,q_1)$ where $q_1=a \exp[i \theta/2]$ and $q_2=b \exp[-i\theta/2]$, where $a,b$ are nonequal amplitudes of the condensates in the dyad and $\theta$ is the phase difference between them. If out of $n$ independent condensation events, the condensate $1$ (say, 'bottom' condensate in a vertically oriented dyad) acquired $q_1$ state $k$ times, then its II becomes $I(n,k)=|k q_1 + (n-k)q_2|^2/n^2=(k^2a^2 + (n-k)^2b^2 +2 k (n-k) a b \cos\theta)n^{-2}.$ The expectation, $\mu$, and the variance, $\sigma^2$, of the distribution of  IIs for condensate '1' can be found directly (see Supp. Inf. for details) from 
$\mu=2^{-n}\sum_{k=0}^n{}^nC_k I(n,k),$ and $
    \sigma^2= 2^{-n}\sum_{k=0}^n{}^nC_k I(n,k)^2 - \mu^2$ and using the standard sums of binomial coefficients $\sum_{k=0}^n {}^nC_k k^p  =\prod_{j=0}^p(n-j)2^{n-p}$ for $p=0,...,4.$ In the limit of large $n$, we get approximately normal distribution with $\mu=(a^2+b^2 + 2 a b \cos\theta)/4$ and $\sigma^2=|b^2-b^4|.$ 
    In Suppl. Inf. we analyze the agreement of IIs and normal distributions. 
    As $a,b$ and $\theta$ are set by the experimental controls, such as the separation of condensates in the dyad and the shape and intensity of the laser pump, the mean and the standard deviation of the distribution are experimentally controlled. 

\begin{figure}[h!]
     \includegraphics[width=\columnwidth]{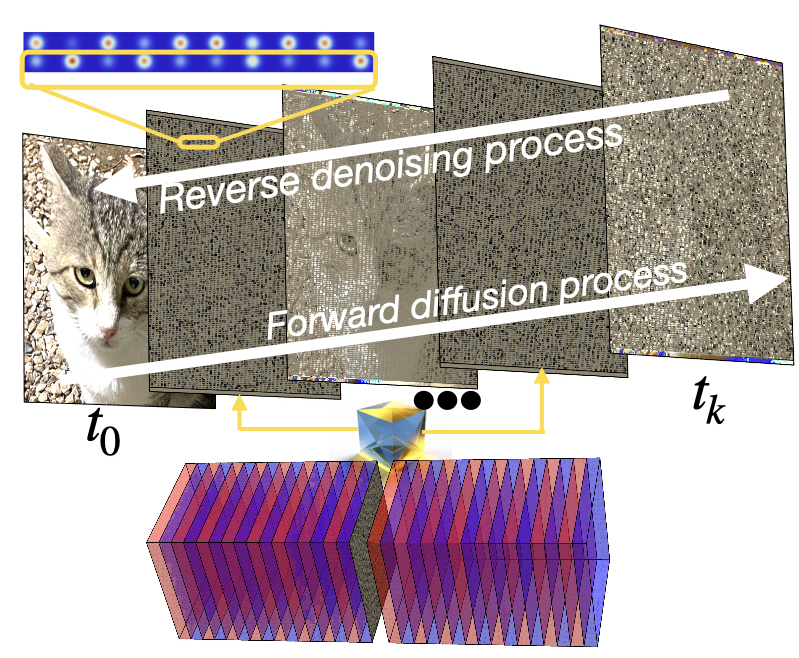}
     \caption{Schematics of the usage of the polariton dyad lattice as optical analog hardware in the generative diffusion process. The lattice of polariton dyads is realised in a microcavity. The emitted light is all-optically guided and superimposed with the pixels of the image at the previous step during the forward diffusion process to transform data into noise gradually. This corresponds to sampling from a normal distribution ${\cal N}(x_t; x_{t-1},(\sigma_t^2 - \sigma_{t-1}^2))$, while the image gives the mean at the previous step. The reverse process learns to transform the noise back into data. }  
    \label{diffusion}
\end{figure}

The forward stage of the generative diffusion process, characterized by the iterative addition of noise to data, poses a substantial computational challenge for electronic digital hardware that can be effectively addressed by integrating the polariton platform as analog hardware. In the forward stage of the generative diffusion process tailored to polariton dyads arranged in a lattice, one condensate in a dyad acts as a source of normally distributed noise; the emitted light can be transmitted all-optically (e.g., guided by optical waveguides)  and superimposed with the image. 
 Adopting these optical technologies in the forward diffusion stage enhances processing speed and energy efficiency and opens up new possibilities for advanced image processing techniques. Figure {\ref{diffusion}} depicts the schematics of the incorporation of the polariton dyad lattices into the generative diffusion process. We estimate the time of the all-optical generation of the Gaussian noise using the typical parameters of a short-lived polariton condensate lattices: $2 cm^2$ samples can accommodate $10^7$ dyads separated by $4\mu$m. With the coherence time of $10$ps the integral light intensity accumulated over typical $10$ns corresponds to sampling for $n=1000$ which gives an accurate fit to the normal distribution (as illustrated in Supp. Inf.). In the all-optical transmission of the integrated light intensity, the time does not depend on the size of the lattice, giving at least three orders of magnitude improvements in speed and energy consumption compared to a GPU \cite{ulhaq2022efficient,michal2023role,stroev2023analog}.

In summary, we have elucidated a macroscopic response leading to density asymmetry in a photonic non-Hermitian dyad and suggested these structures can be used in sensitivity devices, for hRNG, and in generative diffusion models in machine learning, among many other noise-dependent applications. Error correction for unbiased statistics of the dyad orientations is possible with this system. Any bias resulting from asymmetry in the sample can be overcome by modifying the pumping strength at one condensation site. Such systems could sense internal and external irregularities in the sample. Unbiased orientations enable the generation of uniform random number distributions, while controlled bias can be introduced by implementing weak couplings between condensates in different dyads. Integrated intensities of light emanating from dyads are shown to be normally distributed and can be optically transmitted to implement the generative diffusion models of machine learning.   Finally, a significant advantage of constructing this scheme with non-Hermitian optical systems is the speed of random number generation. With potentially hundreds of thousands of asymmetric dyads placed on a chip and the ultra-short picosecond timescales required for laser and exciton-polariton condensates to establish coherence, the sampling of low-level, statistically random signals will occur in parallel and at an ultra-fast timescale. This would bring the random number generation comfortably to the THz regime, bounded only by the signal's conversion speed to the electronic domain.

{\it Acknowledgements.}
A.J. is grateful to Cambridge Australia Scholarships and the Cambridge Trust for entirely funding his Ph.D. N.G.B. thanks the Julian Schwinger Foundation grant JSF-19-02-0005, Horizon Europe/UKRI G123670, and 142568 / Weizmann-UK grant for the financial support.

\section{Supplemental Information}
\subsection{Chain of asymmetric dyads}
\begin{figure}[h!]
     \includegraphics[width=\columnwidth]{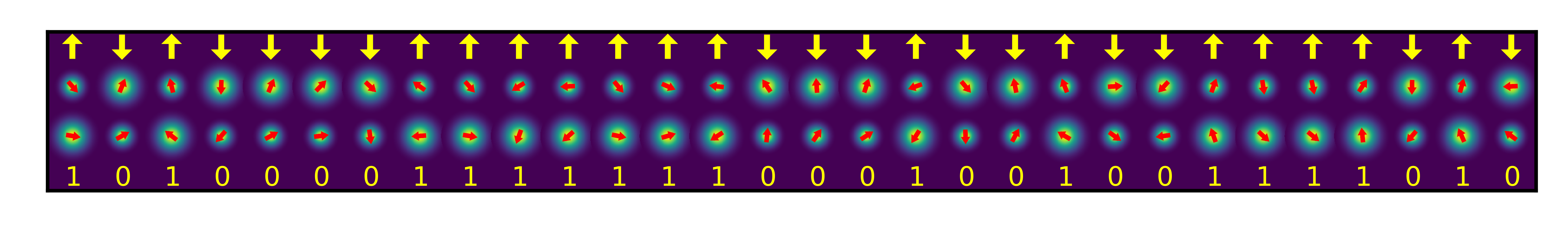}
     \caption{Arrangement of an $N=10$ chain of asymmetric dyads with no couplings between neighboring dyads, representing the integer $716$. Condensate densities are depicted  in a blue (low)-yellow (high) colour scheme while their respective phases are shown with red arrows. The spin representation is shown above the condensates with yellow arrows while the binary representation of each dyad is shown below.}  
    \label{Chain_schematic}
\end{figure}

A chain of $N$ asymmetric dyads has $2^{N}$ possible arrangements, corresponding to the integers ${0,1,2,...,2^{N}-1}$. Fig.~\ref{Chain_schematic} depicts the densities and relative phases of a chain of $10$ condensate dyads  with no couplings between neighboring dyads, representing the integer $716$.   The phase difference within each dyad is not arbitrary - it is a characteristic feature of any particular asymmetric dyad. 

A chain of $N$ dyads, as depicted in Fig.~\ref{Chain_schematic}, is favourable over a single dyad for hRNG, since it would produce a bit rate that is $N$ times faster than the latter. Chains are also advantageous for sensing, since they can survey a wider spatial area than single dyads.

\subsection{Using non-zero inter-dyad couplings to introduce bias}
\begin{figure}[h!]
     \includegraphics[width=\columnwidth]{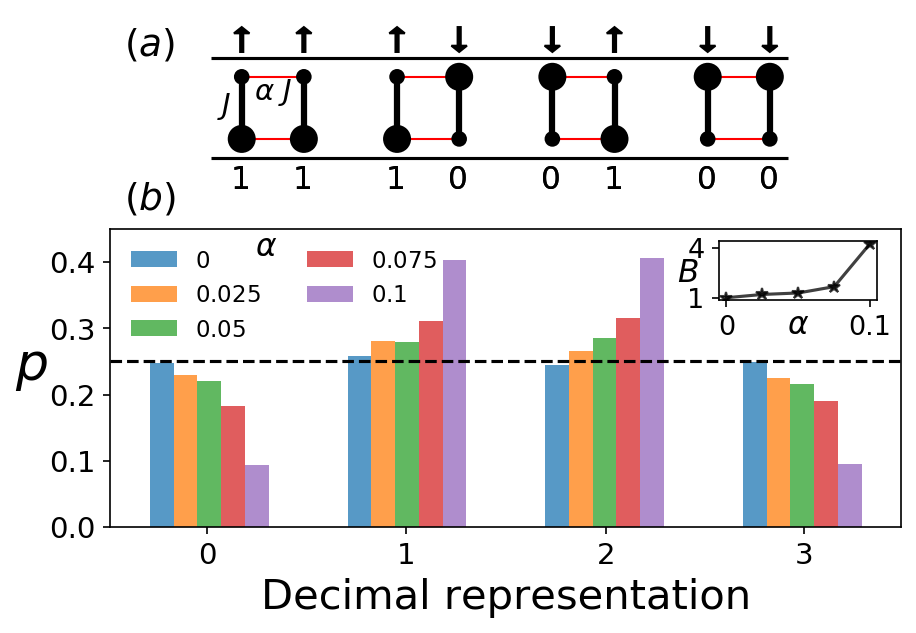}
     \caption{The arrangement of the tetrad system for the four possible final states obtained by numerical integration of Eqs.~(\ref{psi_i_equation}) are shown in (a). Larger (smaller) circles denote higher (lower) density condensates. Thick black  (thin red) lines show the coupling $J$ between condensates within an asymmetric dyad (between condensates in different asymmetric dyads). 
     (b) shows a histogram of the proportion, $p$, of times the system converges to each of the four states for five $\alpha$ values. The unbiased case, $p = 0.25$, is shown with a dashed black line. For each value of $\alpha$, the set of $N$ equations described by Eq.~\ref{psi_i_equation} was solved for $10000$ random initial conditions, with $N=4$ and the $J_{ij}$ coupling matrix elements as prescribed by the diagrams in (a). The inset shows the bias, $B$, as a function of $\alpha$. $J=0.55$, $\gamma=2.8$, $g=0.5$, and $\xi=5/3$.}  
    \label{tetrad_histogram}
\end{figure}

\begin{figure}[h!]
     \includegraphics[width=\columnwidth]{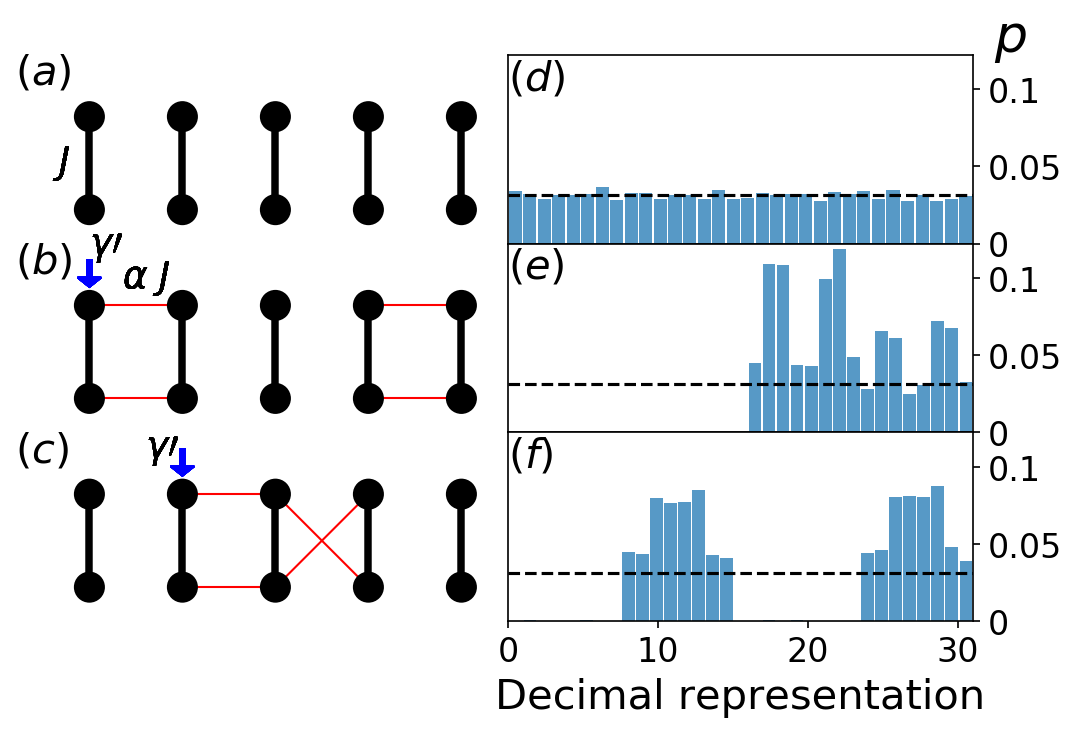}
     \caption{Three  chains of asymmetric dyads [(a),(b),(c)] numerically computed using Eq.~(\ref{psi_i_equation}) for $N=10$ shown alongside their associated probability distributions [(d),(e),(f)]. Distributions were calculated by categorising decimal representations of the final configurations of the system  using $5000$ random initial conditions. The unbiased case, $p = 1/32$, is shown with a dashed black line in each plot. $J_{ij}$ values are prescribed by the corresponding diagrams [(a),(b),(c)]. Circles mark condensate sites, while blue arrows show sites at which an increased pumping strength, $\gamma\prime = 1.05\gamma$, is used. Thick black lines show couplings of strength $J$ while thin red lines show weaker couplings of strength $\alpha J$. (a) shows a chain of non-interacting dyads, producing a uniform random number distribution (d). The biased distributions of (e) and (f) result from the chains in (b) and (c). (b) shows a chain with two lateral couplings with strength $\alpha = 0.1$ and increased pumping strength  at one site. (c) shows a chain with one set of lateral couplings and one set of crossed couplings, both of which have $\alpha = 0.01$, and one site with increased pumping strength. $J=0.55$, $\gamma=2.8$, $g=0.5$, and $\xi=5/3$.}  
    \label{Diagrams_and_histograms}
\end{figure}

Next  we investigate a system of two coupled asymmetric dyads, in which condensates in each dyad are coupled with strength $J$ while adjacent condensates (within different asymmetric dyads) are coupled with strength $\alpha J$, where $\alpha \ll 1$. 
We consider a tetrad configuration in which two asymmetric dyads are weakly coupled to each other. The highest occupation (ground) states of this configuration were investigated in \cite{johnston2021artificial}. Condensates in each asymmetric dyad are coupled with strength $J$ while adjacent condensates (within different asymmetric dyads) are coupled with strength $\alpha J$, where $\alpha \ll 1$. This arrangement is outlined in Fig.~\ref{tetrad_histogram}(a). Fig.~\ref{tetrad_histogram}(a) shows the four possible states ($11$, $10$, $01$, and $00$), while Fig.~\ref{tetrad_histogram}(b) shows the proportion, $p$, of times each state forms over $10000$ trials with random initial conditions, for a range of $\alpha$ values. When $\alpha = 0$ (i.e., two non-interacting dyads) all possible arrangements are equally likely, as expected. 
However, when $\alpha$ is nonzero, there is a bias towards dis-aligned states ($10$ and $01$). The inset of Fig.~\ref{tetrad_histogram}(b) shows that this bias - $B \equiv p_{\rm dis-aligned}/p_{\rm aligned}$ where $p_{\rm dis-aligned}$ ($p_{\rm aligned}$) is the  proportion of dis-aligned (aligned) states - increases nonlinearly with $\alpha$. One can bias the distribution to a significant extent ($B \approx 4$ when $\alpha = 0.1$). The reciprocal bias can be achieved by arranging the sample such that the couplings between dyads cross (creating an ``X"), as shown in Fig.~\ref{Diagrams_and_histograms}(c). 

We show that $\alpha$ can be used to bias the  proportion of aligned to dis-aligned states that form. This can be used to create non-uniform random number distributions. Since the coupling strength between two condensates follows a Bessel function of their separation distance \cite{ohadi2016nontrivial}, adjustments to their spacing can achieve either crossed (i.e., upper condensate within one dyad coupled to lower condensate of adjacent dyad) or lateral couplings (i.e., no crossing). Certain sections of the random number distribution can be eliminated by increasing the pumping strength of one condensate of one or more dyads, making its orientation deterministic. Fig.~\ref{Diagrams_and_histograms}(a) illustrates a uniform distribution created with a chain of non-interacting dyads. Figs.~\ref{Diagrams_and_histograms}(b) and (c) show biased distributions, constructed by modifying couplings and pumping strengths.

\subsection{Derivation of the distribution of the integrated intensity}
We consider the normalized integrated intensity
$$I=\frac{1}{T}\biggl|\int_{t=0}^{t=T} \psi(\tau)\, d\tau\biggr|^2\approx  \frac{1}{n^2}\biggl|\sum_{i=1}^{n=[t/\xi]}\psi_i\biggr|^2,$$
where $\psi_i$ are coherent states of one condensate in the dyad formed in the duration $T$ of  the measurement. For a condensate in an asymmetric dyad two states are possible each with the probability $p=1/2$:  $q_1=a \exp[i \theta/2]$ and $q_2=b \exp[-i\theta/2]$, where $a,b$ are nonequal amplitudes of the condensates in the dyad and $\theta$ is the phase difference between them. If out of $n$ independent condensation events, the  condensate  acquired $q_1$ state $k$ times, then its integrate intensity, denoted as $I(n,k)$ becomes 
\begin{eqnarray}I(n,k)&=&|k q_1 + (n-k)q_2|^2/n^2\nonumber\\
&=&(k^2a^2 + (n-k)^2b^2 +2 k (n-k) a b \cos\theta)n^{-2}.\nonumber
\end{eqnarray}
The expectation, $\mu$ and variance $\sigma^2$ are
\begin{eqnarray}
\mu&=&\sum_{k=0}^n \binom{n}{k} I(n,k) p^k(1-p)^{n-k}, \nonumber \\
\sigma^2&=& \sum_{k=0}^n \binom{n}{k} I(n,k)^2 p^k(1-p)^{n-k} - \mu^2.
\end{eqnarray}
We substitute $p=1/2$ and note that the binomial expansion of $(x+y)^n$ or its derivatives for $x=y=1$ can be used to derive
\begin{eqnarray}
 2^n&=&\sum_{k=0}^n  \binom{n}{k}, \nonumber\\
 n2^{n-1}&=&\sum_{k=0}^n k \binom{n}{k}, \nonumber\\ n(n-1)2^{n-2}&=&\sum_{k=0}^n k^2 \binom{n}{k}, \nonumber\\
 n(n-1)(n-2)2^{n-3}&=&\sum_{k=0}^n k^3 \binom{n}{k}, \nonumber\\  n(n-1)(n-2)(n-3) 2^{n-4}&=&\sum_{k=0}^n k^4 \binom{n}{k}.
\end{eqnarray}
Using these expressions, we evaluate the expectation and variance as 
\begin{eqnarray}
\mu&=&\frac{(n-1) \left(a^2+b^2\right)+2 a b (n+1) \cos (\theta )}{4 n}, \nonumber\\
    \sigma^2&=&\frac{1}{8 n^3}[a^4 \left(-2 n^2+5 n-3\right)-12 a^3 b (n-1) \cos (\theta )\nonumber\\
    &&+2 a^2 b^2 \left(2
   n^2+n-3\right)+4 a^2 b^2 (n-3) \cos ^2(\theta )\nonumber\\
   &&+4 a b^3 (n+3) \cos (\theta )-b^4
   \left(8 n^3+2 n^2+3 n+3\right)+8 b^2 n^3].\nonumber
\end{eqnarray}

In the limit of large $n$, we get  $\mu=(a^2+b^2 + 2 a b \cos\theta)/4$ and $\sigma^2=|b^2-b^4|.$ 

\subsection{Approximating the normal distribution by the integrated intensity}

\begin{figure}[h!]
     \includegraphics[width=\columnwidth]{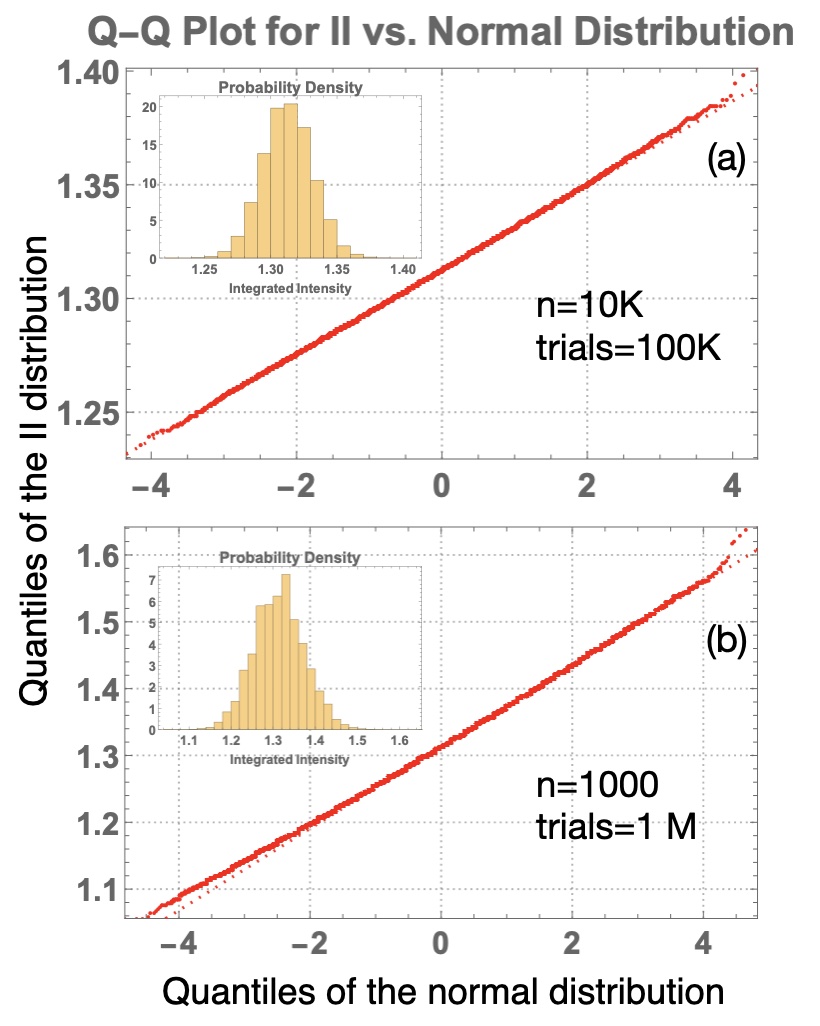}
     \caption{Quantile-Quantile plots that compare the quantiles of the integrated intensity data distribution against the quantiles of a a normal distribution for (a) $n=1000$ and (b) $n=10000$. We took $a=2, b=0.5, \theta=\pi/3$ and generated one million  (a) and $10,000$(b) samples of length $n$. The insets show the histogram of the distributions and confirm the analytical values of the mean and variance, that for these values are $\mu= 1.312$ and the  $\sigma^2=0.187$.}  
    \label{pdf}
\end{figure}

In the previous section we calculated  the mean and variance of the distribution of the integrated intensity for a condensate in the dyad. Here we verify empirically that the distribution approximates the normal distribution with the same mean and variance by  plotting Quantile-Quantile (Q-Q). Figure {\ref{pdf}} compares the quantiles of the integrated intensity data distribution against the quantiles of a a normal distribution. If the data follows the theoretical distribution closely, the points on the Q-Q plot should lie approximately on a straight line which is the case for the integrated intensity distribution. 

\bibliography{references-2,diffusionmodels}

\end{document}